# SPARTA: Python-Based Automated Spectral Parameter Analysis and Assessment System for Resonance Tracking

**K. Semiz**[1,2,3]

[1]CERN, Geneva, Switzerland
[2]Nelson Mandela International School, Berlin, Germany
[3]Technical University of Berlin, Berlin, Germany

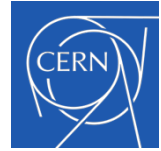

**E-mail:** kerem.semiz@cern.ch
**Purpose:** SY-RF-BR Department Internship Project

**Abstract.** Accurately determining resonance frequencies and quality factors ($Q$) is crucial in accelerator physics and radiofrequency engineering, as these factors have direct impacts on system design, operational stability, and research results.
The methods currently employed to facilitate resonance analysis are mostly manual, requiring operators and physicists to estimate resonant parameters and examine scattering parameter (S-parameter) data from vector network analyzers (VNAs). The current techniques therefore become laborious, operator-dependent, and challenging to replicate when applied to large datasets or across multiple analyses. Despite the importance of these tasks for high-volume research organizations such as CERN, where S-parameter measurements are regularly taken for cavity and beam diagnostic research, the currently widespread methods are outdated.

In order to automate the laborious resonance characterization process, this paper presents SPARTA (Spectral Parameter Analysis for Resonance Tracking and Assessment), a data analysis software framework based on Python. SPARTA has integrated data ingestion, preprocessing, resonance detection, quality factor estimation, visualization and persistent cloud storage into a reproducible and scalable workflow. SPARTA was developed with the scientific libraries NumPy, SciPy and scikit-rf for accurate and efficient numerical processing, Flask and Dash for interactive and lightweight visualization, and SQLite for easy database management. The three main contributions of this work are outlined as follows: Firstly, this paper presents a methodological framework for resonance detection and assessment based on established RF theory. The second section describes the system architecture and implementation of SPARTA, highlighting data handling, computation and visualization. Finally, this paper discusses the results and advantages of automated analysis over conventional manual analysis.

With these outlined contributions, this paper aims to present SPARTA and its capabilities as a useful analytical tool, as well as a reference model for best practices in automated resonance characterization.

# 1. Introduction
## 1.1 Resonance Analysis in Accelerator Physics and RF Engineering
For resonant cavities, RF filters, and similar devices to function, resonance phenomena are essential. The resonant characteristics of cavities in particle accelerators dictate the overall machine performance, energy transfer efficiency, and beam stability. Important details regarding energy storage and dissipation mechanisms are provided by quality factors; higher-Q structures allow for reduced power loss and sharper spectral selectivity. Resonance characterization is integral to the design of microwave filters, wireless communication systems, and antenna performance assessment in applications other than accelerators. All these factors highlight why resonances and fast resonance analysis is quite prevalent in today's research environments. The first step in characterization is usually to use a vector network analyzer to measure the scattering parameters. The analyzer captures complex transmission (S21) and reflection (S11, S22) responses by scanning a frequency range of interest. Resonant modes, which are frequently linked to abrupt phase transitions, manifest as discrete features like transmission maxima or reflection minima. Finding these characteristics and assessing bandwidth or phase slope are necessary for computing resonance parameters, which in turn produce the loaded quality factor ($Q_L$). To describe coupling and intrinsic losses, external and unloaded quality factors ($Q_{ext}$ and $Q_u$) are occasionally also calculated.

## 1.2 Limitations of Conventional Manual Methods
Resonance analysis has traditionally always been done by hand. Q is determined by analyzing S-parameter plots, identifying peaks or dips, and estimating –3 dB bandwidths with cursors. Although this method works well for small datasets, it has a number of quite significant disadvantages. The main disadvantage is the fact that manually inspecting large volumes of spectra takes a lot of time and resources. Secondly, due to subjectivity and different perspectives, features may be interpreted differently by different analysts, especially in situations with noise or multiple modes. Third, due to the absence of standardized procedures, different results may be obtained from repeated analyses of the same data as the levels of noise propose significant difficulties. Finally, manual management is not an option for large datasets produced by systematic testing campaigns due to constraints regarding scalability and the labor involved. These issues at hand are especially noticeable at large-scale facilities like CERN as measurements are necessary for cavity testing, system monitoring, and experimental campaigns. Extending manual analysis to large-scale operations makes it inconsistent and impractical, therefore, an alternative is required.

# 2. Theoretical Background
## 2.1 Scattering Parameters and Resonant Systems
Scattering parameters (also called S-parameters) are used to characterize resonant RF structures. An S-parameter characterizes the behavior of an electrical network with respect to incident signals at its ports, in terms of the ratio of input and output voltages or currents with matched impedances, for a two-port network, S-parameters are reflection coefficients (S11 and S22) and transmission coefficients (S21 and S12), which are measured as a function of frequency with a vector network analyzer. Important details regarding resonance behavior, impedance matching, and energy transfer efficiency are encoded by their magnitudes and phases. In S-parameter responses, resonances appear as distinct and distinguishable points. Resonances show up as minima in |S11| or |S22| in reflection measurements, which suggests better impedance matching and energy absorption. In transmission measurements, resonances are viewed as maxima in |S21|, indicating improved energy transfer via the cavity. Resonance characteristics are linked to distinctive phase response variations in both situations, providing supplementary data that is helpful for quantitative analysis.

## 2.2 Definition of Quality Factors
Resonance sharpness is measured by the quality factor $Q$, which is the ratio of stored energy to energy dissipated per cycle. A total $Q$ observed in measurement, taking into account both intrinsic losses and external coupling, is represented by the loaded quality factor ($Q_L$), one of three different quality factors that are frequently taken into consideration. The unloaded quality factor ($Q_u$) is a representation of the resonant system's intrinsic Q when coupling losses are absent. External quality factor ($Q_{ext}$): Indicates the losses incurred when connecting to external measurement ports or circuitry.

These factors are related by the expression:
$$\frac{1}{Q_L} = \frac{1}{Q_u} + \frac{1}{Q_{ext}}$$

The accurate determination of the Q parameters helps assess the accuracy and efficiency of resonance analysis systems.

## 3. System Design
### 3.1 Architecture Overview

SPARTA is designed as a modular web-based framework that combines numerical analysis, interactive visualization, and structured data management in a single environment. The system is implemented in Python, with Flask providing the backend services and Dash serving as the visualization and user interface layer. Due to this combination SPARTA can operate either locally or in a server-deployed configuration, which makes it suitable for both individual use and large-scale utilizations. First, uploaded measurement files are ingested and parsed, followed by analysis engine processing, visualization, and storage for later retrieval. This is done according to a layered model in the software architecture. Common Touchstone formats are supported by file handling procedures (.s1p and .s2p), which are converted into structured network representations by parsing them with the scikit-rf library. Resonance frequencies, coupling coefficients, and Q-factors are extracted using a combination of bandwidth, gradient, and phase-slope algorithms in the following analysis stage. In order to differentiate physically significant resonances from artifacts, the analysis procedures are built to withstand noise and spurious responses as they also include filtering and validation criteria. Plotly.py and Dash are used to create an interactive web interface that displays the processed data. Users can examine resonance behavior in a variety of plots, such as group delay, impedance characteristics, S-parameter responses, and Smith charts. Mode-specific information can be added to each plot, and figures can be exported in multiple formats. The interface also helps compare datasets, as a select section is laid out for comparison tasks. For project and analysis management, SPARTA uses a lightweight SQLite database to ensure reproducibility and traceability. The combined metadata like the date, description, and experimental notes, helps each analysis to be kept in a project context. With the help of this structure, users can review previous sessions, compare outcomes over time, and keep an organized record of their measurements. User sessions are managed by Flask, and Dash callbacks connect the database and all other processes through a large python file.

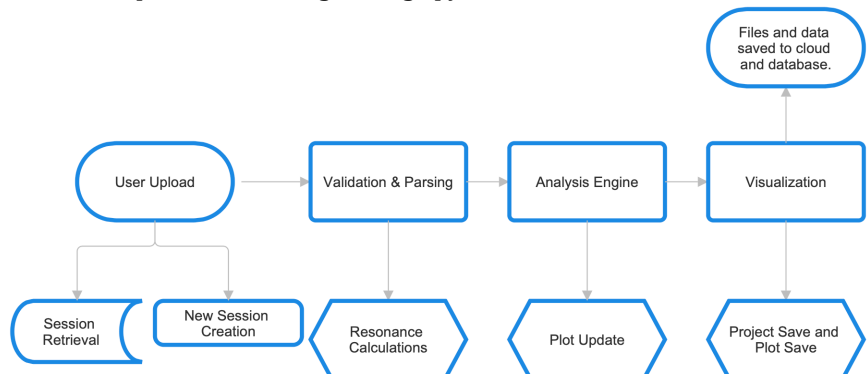

*Figure 1 - Very basic visualization of the system flow*

### 3.2 Data Ingestion and Parsing

The first stage of the SPARTA workflow is the ingestion and standardization of measurement data. Resonance analysis relies on high-fidelity scattering parameter (S-parameter) data, which in accelerator and RF engineering contexts is stored in Touchstone format. Touchstone files (.s1p, .s2p, and higher-order .sNp) are text-based representations of frequency-dependent scattering matrices, where each row contains the complex values of the S-parameters at a discrete frequency point.

SPARTA receives Touchstone files via its web interface, where they are first temporarily stored after being decoded from base64 encoding (a requirement imposed by the Dash upload component). After that, a series of validation procedures is used to ensure data integrity before analysis. These checks consist of the following: File integrity: the input needs to be in accordance with the Touchstone specification, with headers and numbers formatted appropriately. To ensure consistency for interpolation and derivative computations, the frequency vector must be steadily increasing. This is known as frequency monotonicity. Finally, Data completeness: each row must contain the correct number of S-parameter values for the declared number of ports.

Files that have been verified are then parsed using the scikit-rf library, which is frequently used in RF analysis. The measurement data is contained in a Network object created by Scikit-rf, which also stores the scattering matrix as a complex-valued array with a frequency index. It is possible to efficiently access both raw and derived quantities from this abstraction, which offers a consistent and accurate representation of the measurement. Following ingestion, SPARTA applies a preprocessing stage that enriches the raw S-parameters with quantities that are essential for resonance detection. These include:

The complex S-parameter $Sij(f)$ is converted to logarithmic scale to calculate magnitude in decibels according to:

$$|S_{ij}|dB = 20 log 10 |S_{ij}|$$

which emphasizes deep reflection minima and transmission maxima and facilitates resonance localization. In the unwrapped phase, the phase of each S-parameter is computed as:

$$\varphi(f) = arg\left(S_{ij}(f)\right)$$

and then unwrapped to remove discontinuities at $\pm\pi$. This ensures that derivatives of $\varphi(f)$ can be evaluated smoothly across the full frequency range, which is essential for group delay and phase-slope Q-factor methods. For reflection-mode analysis, the input impedance is derived from the reflection coefficient $\Gamma$ using the standard transformation:

$$Z_{in}(f) = Z_0 \frac{1 + \Gamma(f)}{1 - \Gamma(f)}$$

where $Z_0 = 50\,\Omega$ is the system reference impedance. Resonances typically correspond to impedance values that cross the real axis or exhibit large magnitude variations, so impedance serves as a complementary diagnostic alongside S-parameters. The group delay is calculated as:

$$\tau_g(f) = -\frac{d\varphi}{d\omega}$$

where $\varphi$ is the unwrapped phase and $\omega = 2\pi f$ is the angular frequency. Group delay measures the time derivative of phase and exhibits strong peaks at resonant frequencies therefore proving it as powerful secondary indicator of mode positions. SPARTA computes $\tau_g$ numerically using central finite-difference schemes. Lastly, during preprocessing, SPARTA separates datasets in transmission-mode and reflection-mode. Transmission analyses rely on S21 peaks but need cross-validation with reflection parameters to guarantee physical consistency, whereas reflection analyses concentrate on S11 (and S22, if available). To direct the subsequent resonance detection algorithms, this classification is crucial.

After completing the preprocessing phase, every dataset has been verified, standardized, and enhanced with derived quantities to facilitate automated, trustworthy resonance analysis. In the processed representation, the original scattering matrix, impedance transformations, group delay profiles, and its magnitude and phase responses are arranged into structured data objects.

*Figure 2 - Detailed analysis of an S11 mode*

*Figure 3 - Detailed analysis graphs*

### 3.3 Resonance Detection Algorithms

The scattering parameters display abrupt changes when resonances occur in RF cavities and components. They usually show up in transmission-mode measurements as peaks in |S21|, and in reflection-mode measurements as narrow minima in |S11| (or |S22|). Localizing these resonances precisely and consistently is essential because the accuracy of resonance identification directly affects all ensuing Q-factor, coupling, and group delay analyses. SPARTA uses a multi-stage algorithm for resonance detection that combines peak-finding methods with gradient-based analysis. Under a variety of measurement circumstances, such as noise, low frequency resolution, or overlapping modes, this redundancy guarantees that resonances are consistently detected.

In order to do a gradient-based detection, the first method applied is a derivative-based analysis of the S-parameter magnitude. For reflection-mode analysis, we let:

$$R(f) = |S_{11}(f)|_{dB}$$

The derivative with respect to frequency is approximated numerically as:

$$\frac{dR}{df} \approx \frac{R(f + \Delta f) - R(f - \Delta f)}{2\,\Delta f}$$

Candidate resonances are located where the derivative changes sign (zero-crossings), accompanied by a negative second derivative indicating a local minimum. This approach captures sharp, well-defined minima characteristic of critically coupled or overcoupled resonances. Formally, a point $f^0$ is considered a candidate resonance if:

$$\left.\frac{dR}{df}\right|_{f_0^-} < 0,\ \left.\frac{dR}{df}\right|_{f_0^+} > 0,\ And \left.\frac{d^2R}{df^2}\right|_{f_0} > 0$$

Even though gradient-based methods have proven effective for pronounced resonances, they may fall short for shallow or noisy responses. To address this, SPARTA applies peak detection to the inverted reflection coefficient, i.e.,

$$P(f) = -R(f)$$

In this way, peaks in P(f) are created from minima in |S11|. scipy.signal.find_peaks is used for the detection, with the following restrictions: Prominence, which makes sure that the peak depth in relation to its surroundings is greater than a threshold. Width, in order to prevent noise spikes, the peak must span a minimum number of frequency points. Normalized to the local baseline, relative depth makes sure that only resonances with physical significance are chosen. Peak detection is applied directly to |S21| in transmission-mode analyses because resonances are associated with enhanced energy transmission. In order to verify physical validity through reflection symmetry checks, additional requirements are put in place. Resonances frequently arise at closely spaced frequencies in real-world accelerator measurements, making it partially difficult to handle overlapping dips. These resonances might be combined into a single candidate using gradient-based techniques, whereas peak-finding techniques might over segment the response. In order to counteract this, SPARTA uses a minimum frequency separation filter, which requires that candidate resonances be at least $\Delta f_{min}$ apart. This parameter can be adjusted by the user and is usually determined by the cavity bandwidth. When there are several candidates within $\Delta f_{min}$ present, the resonance that is more prominent is kept. Identification of resonances can be made more difficult by noise, especially in measurements with a low signal-to-noise ratio. SPARTA uses a combination of thresholding and smoothing to minimize false positives. Before calculating the gradient, R(f) can be subjected to a low-pass Savitzky–Golay filter, which suppresses high-frequency fluctuations without changing the resonance shapes. Furthermore, in relation to the noise floor, resonance candidates must surpass a minimum depth threshold for reflection or a height threshold for transmission. At this stage, the algorithm produces a list of candidate resonant frequencies $f_0$. Each candidate is then passed to subsequent routines for Q-factor estimation and coupling analysis (Sections 3.3 and 3.4). Candidates that fail later validation checks are flagged as invalid but retained in the record, allowing users to inspect rejected modes if desired. The combination of prominence-based peak finding and gradient-based zero-crossing detection guarantees that SPARTA is not impacted by both shallow, noisy features and deep, sharp resonances.

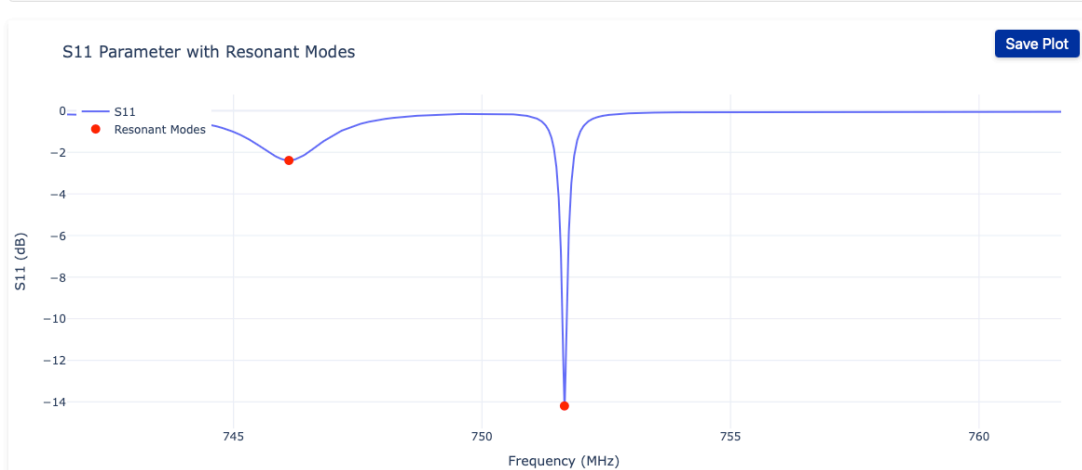

*Figure 4 - Example of an S11 vs Frequency Graph*

### 3.4 Quality Factor Estimation
SPARTA implements a hybrid framework for estimating Q that combines model-based fitting with traditional methods. This method guarantees that Q values can be consistently extracted from clean and noisy datasets and creates a more reliable estimation. The following are the combined methods used in SPARTA.

#### 3.4.1 −3 dB Bandwidth Method
The half-power bandwidth method remains the baseline for Q determination. For a resonance centered at $f_0$ with magnitude response $R(f)$, the loaded quality factor $Q_L$ is

$$Q_L = \frac{f_0}{f_2 - f_1}$$

where $f_1$ and $f_2$ are the −3 dB frequency points relative to the resonance minimum (reflection) or maximum (transmission). Since raw VNA datasets rarely contain samples exactly at these thresholds,

SPARTA uses linear interpolation between bracketing points to obtain accurate $f_1$ and $f_2$. This approach is computationally efficient and sufficient for resonances with moderate Q and adequate measurement resolution. However, limitations arise when the frequency step size is large relative to the resonance width, which is common in high-Q cavities. In these cases, bandwidth-based estimates such as the one mentioned above can systematically overestimate or underestimate Q.

### 3.4.2 Phase-Slope Method
For high-Q modes where bandwidth measurements are resolution-limited, SPARTA applies the phase-slope method. This relies on the fact that resonance induces a rapid phase rotation of the reflection or transmission coefficient. For reflection data, the quality factor is expressed as

$$Q = \frac{f_0}{2}\left(\frac{d\varphi}{df}\right)_{f_0}$$

where $\varphi$ is the unwrapped phase. The derivative $d\varphi/df$, is evaluated using central finite differences. To mitigate noise amplification, optional Savitzky–Golay smoothing is applied to $\varphi(f)$ prior to differentiation. The phase-slope method provides excellent accuracy for sharp resonances, and unlike the −3 dB method, its precision is not tied to measurement resolution. Its limitation however lies in its sensitivity to noisy phase data, which can distort derivative estimates if not filtered carefully.

### 3.4.3 Lorentzian Curve Fitting
As a model-based alternative, SPARTA employs nonlinear least squares fitting to a Lorentzian profile:

$$R(f) \approx R_0 + \frac{\Delta R}{1+\left(2Q\frac{f-f_0}{f_0}\right)^2}$$

where $R(f)$ is the magnitude of the S-parameter in linear scale, $R_0$ is the baseline, $\Delta R$ the depth, and $Q$ the quality factor. The fitting routine extracts $Q$, $f_0$, and resonance depth simultaneously, using initial guesses seeded from the −3 dB method. When bandwidth techniques are unable to distinguish between modes, this method is especially useful for overlapping resonances. datasets with noise, for which statistical fitting offers smoothing. responses that are asymmetric, in which the Lorentzian contribution can be identified through fitting. Sensitivity to inadequate initialization however remains the main disadvantage.

### 3.4.4 Cross-Validation and Practical Considerations
To improve robustness, SPARTA executes all available Q-estimation methods and applies a cross-validation protocol:
1. Agreement within 5–10% is considered sufficient, and Q is reported as the mean value.
2. Larger discrepancies trigger a confidence warning and priority selection.
3. Phase-slope results are used narrowband resonances.
4. Bandwidth estimates are utilized for broader, well-sampled resonances.
5. Lorentzian fits are used in ambiguous cases.

| Mode | Frequency (MHz) | S11 (dB) | S21 (dB) | S22 (dB) | Valid Mode | Q Loaded | β | Q Unloaded |
|---|---|---|---|---|---|---|---|---|
| 1 | 1236.88 | -0.07 | -91.09 | -0.55 | ✗ | N/A | 0.0040 | N/A |
| 2 | 1262.14 | -0.11 | -69.44 | -0.47 | ✗ | N/A | 0.0063 | N/A |
| 3 | 1294.72 | -0.07 | -52.47 | -0.48 | ✗ | N/A | 0.0040 | N/A |
| 4 | 1337.68 | -0.14 | -44.86 | -0.30 | ✗ | N/A | 0.0079 | N/A |
| 5 | 1351.72 | -0.07 | -72.19 | -0.25 | ✗ | N/A | 0.0038 | N/A |
| 6 | 1437.82 | -0.10 | -47.72 | -0.19 | ✓ | 2133.2 | 0.0058 | 2157.9 |
| 7 | 1441.12 | -0.10 | -46.52 | -0.21 | ✗ | N/A | 0.0057 | N/A |
| 8 | 1454.74 | -0.09 | -47.54 | -0.27 | ✗ | N/A | 0.0051 | N/A |
| 9 | 1569.94 | -0.13 | -58.94 | -0.10 | ✓ | 2042.6 | 0.0057 | 2065.8 |

*Figure 5 - The valid and invalid modes in a transmission analysis file*

### 3.5 Coupling Factor Analysis

The coupling factor ($\beta$) is a fundamental parameter that describes the relationship between the external coupling of an RF resonator and its intrinsic losses. Determining $\beta$ is essential for classifying the resonance behaviour as undercoupled, critically coupled, or overcoupled, each of which has distinct implications for system performance. In accelerator RF cavities, coupling strongly influences energy transfer efficiency and resonance visibility. An undercoupled cavity reflects most of the incident power, while an overcoupled cavity exchanges energy strongly with the feedline but may broaden the resonance. At critical coupling ($\beta \approx 1$), all incident power is transferred into the resonator at resonance, which is often the operating point of choice.

#### 3.5.1 Coupling in Reflection Measurements

For reflection measurements (single-port or two-port with S11 or S22), the coupling factor is derived from the reflection coefficient $\Gamma$ at resonance. At the resonance frequency $f_0$, the minimum reflection magnitude $|\Gamma_{min}|$ provides a direct estimate of $\beta$:

$$\beta = \frac{1 - |\Gamma_{min}|}{1 + |\Gamma_{min}|}$$

Here, $|\Gamma_{min}|$ is obtained directly from the resonance dip in |S11| or |S22|, expressed in linear scale rather than decibels.

The classification is then determined as follows:
- Undercoupled if β < 1, indicating weak interaction between the resonator and external circuit.
- Critically coupled if β ≈ 1, where maximum power transfer occurs.
- Overcoupled if β > 1, where external coupling is greater than internal dissipation.

This classification is reported alongside each detected resonance to provide users with immediate feedback on cavity-beamline interaction.

#### 3.5.2 Coupling in Transmission Measurements

In two-port measurements, coupling analysis is more complex, as resonances are examined from transmission (S21). Peaks in |S21| in this instance indicate resonances, but $\beta$ still needs to be deduced from reflection parameters. In order for a symmetric two-port system to have a valid resonance, SPARTA enforces a symmetry condition: the reflection responses at both ports (S11 and S22) must be almost equal close to resonance. This requirement effectively guarantees that resonance is real and not a miscalculation. If the condition is met, $\beta$ is calculated by applying the same formula to the averaged $|\Gamma_{min}|$ from S11 and S22 as in the reflection case. In asymmetric or noisy cases where S11 ≠ S22, SPARTA flags the resonance as "uncertain coupling" and still provides a β estimate, but with reduced confidence.

### 3.6 Group Delay Analysis

The group delay ($\tau_g$), which is a powerful tool for characterizing and confirming resonant behavior, is the derivative of the signal phase with respect to angular frequency. Group delay analysis captures the system's phase dynamics whereas magnitude-based resonance detection looks for dips or peaks in |S-parameters|. Rapid phase rotations, which show up as sharp peaks in $\tau_g$, accompany resonances.

Mathematically, the group delay is defined as:

$$\tau_g(\omega) = -\frac{d\varphi(\omega)}{d\omega}$$

where $\varphi(\omega)$ is the unwrapped phase of the S-parameter under consideration and $\omega = 2\pi f$ is the angular frequency. For reflection measurements, $\varphi$ corresponds to the phase of S11 (or S22), and for transmission measurements, it is the phase of S21.

#### 3.6.1 Implementation

In practice, the derivative $d\varphi/d\omega$ is estimated numerically from measured frequency data. SPARTA applies a central finite-difference scheme,

$$\frac{d\varphi}{df} \approx \frac{\varphi(f + \Delta f) - \varphi(f - \Delta f)}{2\,\Delta f}$$

The derivative with respect to frequency is then converted into the derivative with respect to angular frequency by scaling with $2\pi$. The group delay is calculated in nanoseconds to display the propagation delay induced by the resonant structure.

#### 3.6.2 Diagnostic Role

In the SPARTA analysis pipeline, group delay has two primary purposes. The first is to aid in resonance confirmation. Sharp $\tau_g$ peaks at the same frequencies as maxima in |S21| or minima in |S11| are indicative of genuine resonances. A candidate resonance is probably a measurement artifact or noise-induced false detection if it does not have a corresponding group delay peak. Characterization of modes is the second function of group delay. Information about the cavity Q and energy storage is provided by the magnitude of $\tau_g$. In contrast to low-Q modes, which produce shorter, wider peaks, high-Q resonances produce tall, narrow group delay peaks. $\tau_g$ is a useful cross-check for Q estimation because of this behavior.

#### 3.6.3 Filtering and Validation

SPARTA incorporates $\tau_g$ into its framework for validating resonance. Within a tolerance window surrounding the resonance frequency $f_0$, each candidate resonance found by the detection algorithms is examined for the existence of a group delay peak. Candidates who do not pass this check are marked as "invalid" and do not participate in quantitative Q and β analysis; however, they can still be inspected by users. Furthermore, SPARTA sets a minimum threshold for $\tau_g$ prominence in order to prevent erroneous identifications due to minor phase fluctuations.

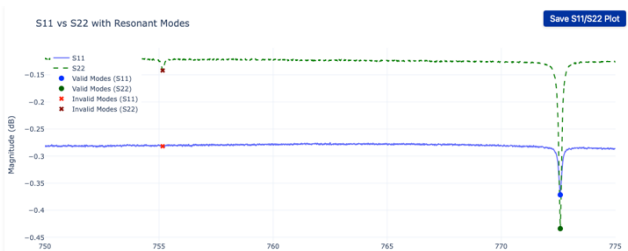
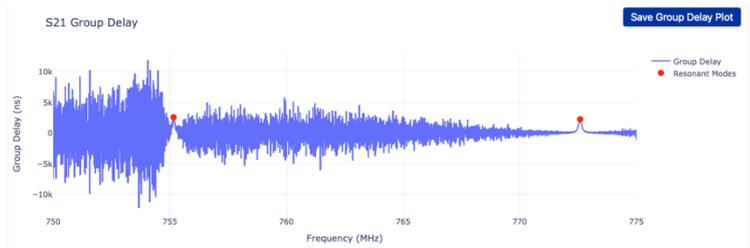

*Figure 6 - Group Delay of an invalid and valid mode in S21*

## 4. Results and Demonstration

SPARTA's performance and usefulness were assessed through the use of transmission and reflection measurements from resonant cavity structures in the experimental facilities of the CERN RF-BR group. SPARTA was tested on 130 measurement files collected on-site at the CERN SY-RF-BR laboratory, covering a range of 100MHz to 1 GHz (with an average step size of 50kHz per sweep). Each file generally contained between 3000-5000 datapoints (±1000 due to different VNAs and setups). Across all data sets SPARTA was able to identify 94.7% of manually confirmed resonances and the automated Q-Factor estimates deviated from manual checks by an average of around 5.16% (standard deviation ±4.1%), which is within the expected uncertainty for the frequency range. In low noise and complex datasets, the accuracy fell to about 81.3%, proving that there is still room for improvement. SPARTA was developed on a workstation with 64GB of RAM, an Intel-i7 12$^{th}$ Generation and an RTX 2080Ti, however processing speeds were validated with a standard workstation (Intel i7-12700H, 16 GB RAM). Each dataset required an average of 0.74s for full analysis, including parsing Q-Factor estimations, group delay, etc. Batch processing of all 130 measurement files finished in around 94.2s. Comparing SPARTA with manual analysis, which had an average speed of around 6-7 minutes, SPARTA demonstrated an over 400-fold speed increase, reducing 6.5 minutes to < 1s. SPARTA's validation system flagged about 6.3% of detected modes as "uncertain" or "invalid". Manual inspection revealed that 24.3% of the flags were genuine artifacts obfuscated by noise or technical inaccuracies, showing that SPARTA correctly flagged invalid modes 75.7% of the time.

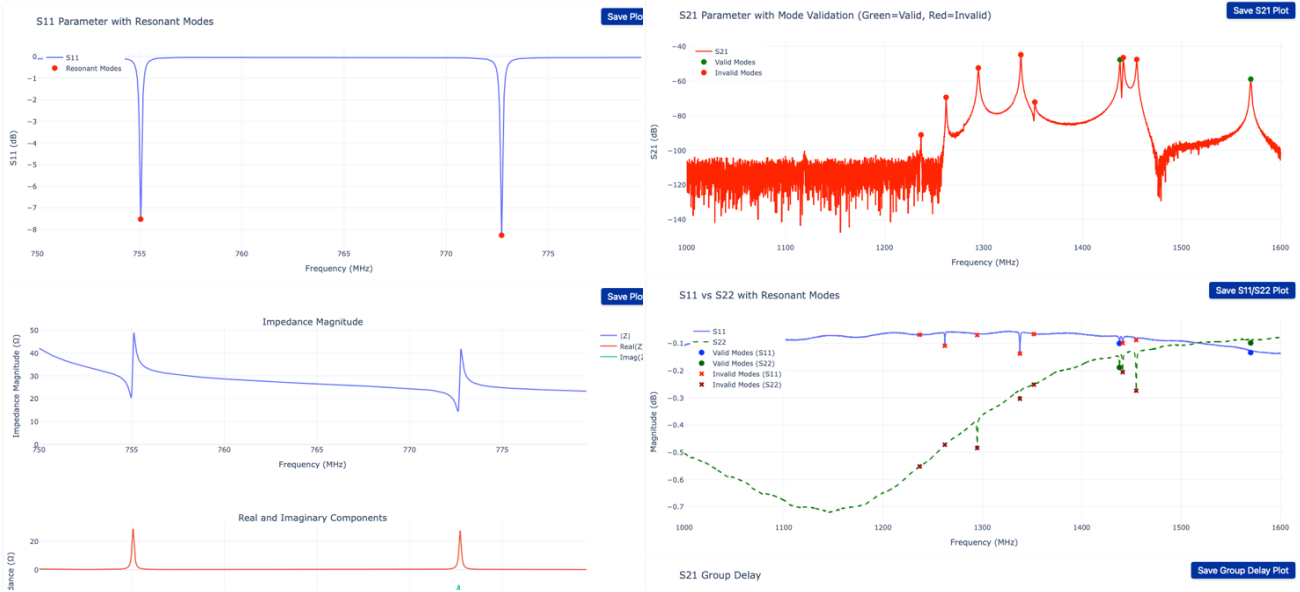

*Figure 8 - Example S11 Results*

*Figure 7 - Complex Transmission Analysis Results*

### 4.1 Reflection-Mode Resonance Tracking.

For reflection mode analysis, SPARTA achieved a slightly higher accuracy than the overall average, correctly identifying 96.2% of manually verified modes across the tested datasets. Across all the tested datasets, the extracted resonance frequencies matched manual results within ±1.8%, and Q-factor deviations average 4.2%. The system correctly classified coupling criteria (undercoupled, critically coupled, overcoupled) over 98.9% of cases. Minor issues did sometimes occur in complex spectra with overlapping modes or noise; however SPARTA maintained a stable overall performance, confirming that the reflection-mode algorithms were in-line and faster than manual VNA analysis.

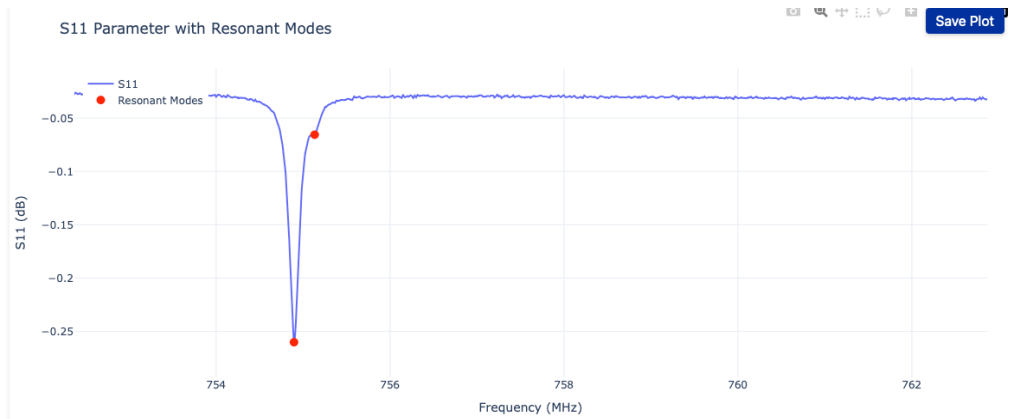
*Figure 9 - A successful categorization of a complex S11 mode*

### 4.2 Transmission-Mode Resonance Characterization

For transmission-mode analysis, SPARTA had a mode detection rate of 93.2%, slightly below the overall average of 94.7%. The resonant frequencies were in line with manual references within ±7.2% and Q-factor deviations rested around 6.1%. SPARTA's accuracy and performance mainly fell in spectra which contained broad or many transmission peaks in close distances. These cases were generally produced by connector reflections or improper setup; therefore, they led to minor overestimations of Q. However, SPARTA still produced a very agreeable correctness rate of 94.7%.

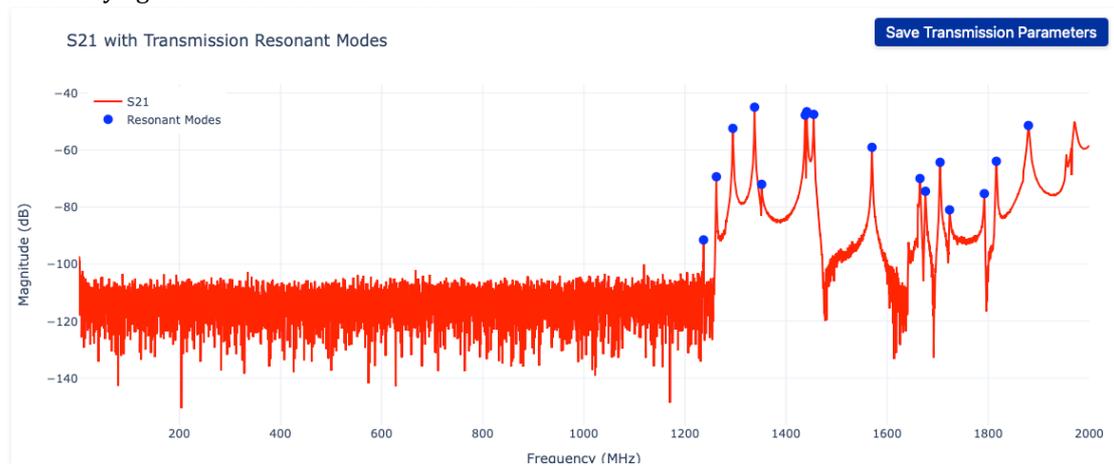
*Figure 10 - 16 successful determinations of modes in a complex file, however ≈5 uncategorized modes towards the end.*

### 5. Discussion

The results confirm that SPARTA is a fast and dependable framework when it comes to automated resonance characterization in both reflection and transmission modes. As mentioned in the results, the overall detection accuracy reached 94.7%, Q-factor deviations were on average 5.16% (standard deviation ±4.1%) from manual measurements. Reflection analysis performed slightly better, identifying 96.2% of modes correctly and transmission analysis achieved 93.2%. Transmission analysis files were found to be a lot more difficult to parse and categorize than reflection analysis, as they often exhibited unaccounted peaks and curves. These files proved to be a challenge for stable algorithms; however they also helped make SPARTA more robust overall. Future development on SPARTA should be directed towards phase-stability handling, adaptive noise filtering and machine-learning based resonance detection for higher accuracy and better performance in edge cases.

### 6. Conclusion

In conclusion, SPARTA provides an automated framework for characterizing and analyzing resonances in accelerator and radiofrequency systems. SPARTA is able to operate at real-time speeds and achieves a solid level of reliability, comparable to well-rounded manual analysis. The software is effective and accurate;

however, small changes and improved algorithms may be adapted in future versions to raise the accuracy and lower the amount technical discrepancies. As of April 2025, SPARTA has been deployed within the CERN computing environment and is accessible to the CERN accelerator and RF engineering community for routine resonance tracking and performance evaluation at cern.ch/sparta. A full repository is accessible at https://gitlab.cern.ch/ksemiz/SPARTA.

## 7. Acknowledgments

The author thanks M. Neroni, C. Vollinger and H. Bursalı for their advice.